\documentclass[aps,prc,twocolumn,amsfonts,showpacs,floats,superscriptaddress,preprintnumbers,byrevtex,nofootinbib]{revtex4}
\usepackage{epsfig}
\newcommand{\etal}{\emph{et al.}}

\newcommand{\nuc}[2]{$^{#1}${#2}}
\begin{document}
\title{Systematics of Fission Barriers in Superheavy Elements}
\author{T. B\"urvenich}
\affiliation{Theoretical Division,
             Los Alamos National Laboratory,
	     Los Alamos, New Mexico 87545}
\author{M. Bender}
\affiliation{Service de Physique Nucl{\'e}aire Th{\'e}orique,
             Universit{\'e} Libre de Bruxelles,
             CP 229, B--1050 Brussels, Belgium}
\author{J. A. Maruhn}
\affiliation{Institut f\"ur Theoretische Physik,
             Universit\"at Frankfurt,
             Robert-Mayer-Strasse 8--10,
             D--60325 Frankfurt am Main, Germany}
\affiliation{Joint Institute for Heavy--Ion Research,
             Oak Ridge National Laboratory,
             P. O. Box 2008, Oak Ridge, Tennessee 37831}
\author{P.--G. Reinhard}
\affiliation{Joint Institute for Heavy--Ion Research,
             Oak Ridge National Laboratory,
             P. O. Box 2008, Oak Ridge, Tennessee 37831}
\affiliation{Institut f\"ur Theoretische Physik II,
             Universit\"at Erlangen--N\"urnberg,
             Staudtstrasse 7, D--91058 Erlangen, Germany}
\preprint{LA-UR-03-104}
\date{February 3 2004}
%
%=======================================================================
%
\begin{abstract}
We investigate the systematics of fission barriers in superheavy elements
in the range \mbox{$Z=108$}--120 and \mbox{$N=166$}--182. Results 
from two self-consistent models for nuclear structure, the relativistic 
mean-field (RMF) model as well as the non-relativistic Skyrme-Hartree-Fock 
approach are compared and discussed. We restrict ourselves to 
axially symmetric shapes, which provides an upper bound on static 
fission barriers. We benchmark the predictive power of the models
examining the barriers and fission isomers of selected heavy actinide 
nuclei for which data are available.
For both actinides and superheavy nuclei, the RMF model systematically 
predicts lower barriers than most Skyrme interactions. In particular the 
fission isomers are predicted too low by the RMF, which casts some 
doubt on recent predictions about superdeformed ground states 
of some superheavy nuclei.
For the superheavy nuclei under investigation, fission barriers 
drop to small values around \mbox{$Z=110$}, \mbox{$N=180$} and 
increase again for heavier systems. 
For most of the forces, there is no fission isomer for superheavy nuclei,
as superdeformed states are in most cases found to be unstable with 
respect to octupole distortions.
\end{abstract}
\maketitle
%
%=======================================================================
%
\section{Introduction}
The search for superheavy elements (SHE) has made exciting progress in
the last few years \cite{Hof98a,Hof00a,Arm00b}. New, often 
more neutron-rich, isotopes of elements \mbox{$Z=108$} \cite{exp108}, 
\mbox{$Z=110$} \cite{exp110}, and \mbox{$Z=112$} \cite{exp112} have been
reported as well as the synthesis of the new elements
\mbox{$Z=114$} \cite{exp114} and \mbox{$Z=116$} \cite{exp116}.
At the same time earlier experiments were confirmed and
more data for already existing isotopes collected \cite{GSIexp112}.
Together with the $\alpha$-decay products of the newly synthesized nuclei 
the known region of superheavy elements has grown substantially.

SHE are by definition those nuclei at the upper end of the chart of
nuclei where quantum-mechanical shell effects reverse the trend of 
decreasing -- and, for the heavier ones, practically vanishing -- 
liquid-drop fission barriers to produce significant stabilization.
The lifetimes of the recently found SHE are many orders of magnitude
smaller than the early optimistic estimates \cite{Nix69,Gru69}.
Additionally the systematics of fusion cross sections suggests that
the extension of the chart of nuclides might be limited by the
production mechanism of SHE, not their decay \cite{Arm00b}. With
recent experiments heading for unknown territory, theory has to
provide reliable predictions for the stability and the most accessible
regions in the landscape of nuclides. A crucial feature is here
spontaneous fission which is characterized by the fission barrier.

Not too much is known experimentally on the fission barriers of
transfermium nuclei. Although their height is not known, barriers 
for $^{252}$No and $^{254}$No are high enough to stabilize these
isotopes against fission up to angular momentum 20 \cite{Rei99aE}. 
An analysis of all available data for fusion and fission of 
$^{292}$112, $^{292}$114 and $^{296}$116 was given recently in 
Ref.\ \cite{Itk02a}. Surprisingly the barrier heights deduced
for these very heavy nuclei are similar, or even slightly larger, 
than the ones of actinide nuclei in the $^{240}$Pu region.

For some other superheavy nuclei it is known that the barrier is relatively 
small from the simple fact that fission is their preferred decay channel.
This leads to another aspect of the stability against fission: 
the experimental identification of new superheavy nuclides is much 
simpler when $\alpha$ decay is the dominating decay channel. All 
recent new decay chains from Dubna end in a region of fissioning 
nuclei which prevents an extension of the known region of SHE to 
the ``southeast'' with current experimental techniques.

The calculation of fission half-lives is a very demanding task,
which became clear quite early \cite{Hil53}. 
First explorations of the potential energy surfaces of transfermium 
nuclei demonstrated already that triaxial and 
reflection-asymmetric degrees of freedom often greatly reduce the 
fission barrier. The collective mass provides the metric for the 
dynamical calculation of the fission half-lives. There is no published 
work so far that considers all ingredients using self-consistent models.
First ambitious steps in that direction were taken for the calculation
of the decay of the fission isomer into the ground state \cite{Kri94a}.
The published fission half-lives from macroscopic-microscopic mean-field 
models also simplify the task considerably by restricting shape-degrees
of freedom to axially symmetric ones and using a phenomenological
parameterization of inertia parameters.
The detailed potential energy landscape is also an ingredient for
estimates on the fusion cross section, although somewhat different
paths have to be considered asymptotically, see e.g.\ 
Refs.\ \cite{Row91a} and further references therein.

It is the aim of this paper to analyze the extrapolation of self-consistent 
mean-field models, namely the Skyrme-Hartree-Fock (SHF) approach and the 
relativistic mean-field (RMF) model, concerning large-amplitude deformation 
properties of SHE. To that end, we present a systematic survey of 
fission paths and barriers for a broad range of SHE, scanning the 
$\alpha$-decay chains of even-even nuclei that are accessible with 
the current experimental techniques. The nuclei considered in this 
study cover the proton numbers \mbox{$108 \leq Z \leq 120$} and 
the corresponding neutron numbers \mbox{$166 \leq N \leq 182$}.
We first benchmark our models by calculating axial fission barriers
and isomeric states for heavy actinide nuclei for which experimental
data are available, namely isotopes with proton numbers ranging from
\mbox{$Z=90$} to 98. 

Section \ref{sect:early} summarizes the most important results of 
earlier studies relevant for our calculations.
Section \ref{sec:theory} explains the theoretical and numerical
methods which are used in our study.
In section \ref{sect:actinides} the results for actinides are discussed.
Section \ref{sec:lands} presents detailed results for deformation
energy curves in all considered superheavy nuclei and for all the 
different forces. Section \ref{sec:chara} discusses the result in 
terms of key quantities such as deformation energies and barriers.
Finally, section \ref{sec:search} attempts to identify the underlying
reasons for the different predictions among the forces and models.
%
%=====================================================================
%
\section{Earlier Calculations}
\label{sect:early}
Estimates of the stability of SHE against fission have a long history. 
First explorations of the potential landscapes were made with
phenomenological corrections for shell effects \cite{Mye66a,Gru69}, 
even before Strutinsky introduced the microscopic-macroscopic method
\cite{Str67a}, which was then immediately applied to large-scale
calculations of fission barriers of heavy and superheavy nuclei,
see e.g.\ Ref.\ \cite{Bol71a}. 
In this framework, the deformation energy is minimized
in a limited space of shape parameters. It soon became clear that
reflection-asymmetric \cite{earlyasym} and triaxial \cite{earlytriax}
shape degrees of freedom have to be included. Complementing the
collective potential energy surfaces by
mass parameters enabled the calculation of fission
half-lives of heavy and superheavy nuclei, with phenomenological mass
parameters \cite{Nix69,earlyhalflives} as well as with 
microscopically computed cranking masses \cite{Pau71}.
Studies along that line are continued until today with improved
parameterizations of the mic-mac model, either the finite-range
droplet (FRDM) plus folded-Yukawa single-particle potential model and
phenomenological masses \cite{Mol87a} or the Yukawa-plus-exponential
(YPE) macroscopic plus a Woods-Saxon microscopic model and cranking
masses \cite{Pom81a,Smo97a,Ghe99a}.  Nearly all large-scale
calculations of fission lifetimes, however, consider axially symmetric
shapes only (as we will do). A recent exception is presented in Ref.\
\cite{Ghe99a} where triaxial shapes are taken into account, however,
at the price of a reduced number of shape degrees of freedom 
in other places.

There do exist also systematic calculations of fission barriers
where the macroscopic part of the energy is calculated within 
the semi-classical Thomas-Fermi approximation \cite{Mye99a}.
One step further toward self-consistency is the extended Thomas-Fermi 
Strutinsky-integral (ETFSI) approach where the same microscopic 
Skyrme force is used to calculate the macroscopic part of the 
binding energy and to determine the single-particle spectra for 
the calculation of the shell correction.
For a large-scale survey of the axially and reflection-symmetric potential
energy surfaces of heavy and superheavy nuclei in the ETFSI approach
using the Skyrme interaction SkSC4 see Refs.\ \cite{Mam98a,Mam01a}.
Results allowing also for triaxial degrees of freedom are presented 
for selected nuclei in Ref.\ \cite{Dut00a}.

To determine the fission path in microscopic-macroscopic, 
Thomas-Fermi, and ETFSI calculations, the total energy is 
minimized with respect to parameters of the nuclear shape. 
There are numerous parameterizations to be found in the literature, 
which differ by emphasizing either high-order multipoles at small 
deformation, or the fragment deformation
of two-center-type configurations at large deformations, see Ref.\
\cite{Has88a} for an overview. While the latter are important for the
proper description of the saddle-point of actinide nuclei (which is 
located at very large $\beta_2$), the first might be better suited 
for superheavy nuclei where the saddle point is located at rather 
small $\beta_2$. The number of shape degrees of freedom in actual 
calculations does rarely exceed five, in most cases it is even smaller.

This restriction does not exist in the framework of 
self-consistent models. Besides very general spatial symmetries that are 
imposed (e.g.\ axiality, reflection symmetry, or triaxiality) there are
no further assumptions made on the nuclear density distribution.
As this makes self-consistent calculations more costly in terms of 
computational time, there is much less published work employing 
self-consistent models so far.

A systematic study of the deformation energy of superheavy nuclei
along the valley of $\beta$ stability in the region
\mbox{$100 \leq Z \leq 128$} and \mbox{$150 \leq N \leq 218$}
in HFB calculations with the Gogny force D1s under restriction to
axially and reflection symmetric shapes was presented in Ref.\
\cite{Ber96a}. Potential energy surfaces of selected heavier
nuclei are presented in Ref.\ \cite{Dec03a}.
The full potential energy surface in the $\beta$--$\gamma$ plane of a
few selected nuclei as resulting from SHF calculations in a triaxial
representation is presented in Ref.\ \cite{Cwi96a}.  This investigation
points out the importance of triaxial shapes at small
deformations \mbox{$\beta_2 < 0.6$}. The fission barrier of some
superheavy nuclei is reduced to half its value when relaxing the
constraint on axial symmetry and going through triaxial paths.
The fission path returns to axially symmetric shapes at larger 
deformations. But here it is necessary to allow for reflection 
asymmetric shapes to accommodate the usually asymmetric fission.
A first exploration of asymmetric shapes of SHE within self-consistent
SHF and RMF models was presented in Ref.\ \cite{Ben98a}. 
For many superheavy systems, reflection-asymmetric shapes lower the 
fission path at large prolate deformation \mbox{$\beta_2 > 0.6$}, 
and remove in most cases the outer barrier known from actinide
nuclei (and persisting for superheavy nuclei when 
considering reflection-symmetric shapes only).
%
%===========================================================================
%
\section{Theoretical Framework}
\label{sec:theory}
%
%---------------------------------------------------------------------------
%
\subsection{Effective Interactions}
We explore the potential landscapes using two widely used
self-consistent mean-field models, namely the non-relativistic 
Skyrme-Hartree-Fock (SHF) method as well as the relativistic mean-field 
(RMF) approach \cite{RMP}. There exists a great variety of parameterizations 
for both models which often differ when extrapolated.
It is long known that different parameterizations of a
self-consistent model are not equivalent for the calculation of
fission barriers of actinide nuclei, see Ref.\ \cite{Dut80a,Bra85aR} 
for a comparison of early Skyrme forces and \cite{Rut95a} for 
a comparison of RMF forces. Comparisons between SHF and RMF hint
at genuine model differences \cite{Rut95a,Ben00b}. Relevant
for our study is also that there exist conflicting predictions 
for the location of the spherical magic numbers in SHE, 
see Ref.\ \cite{Rut97a,Ben99a}, which can be expected to be reflected 
in the structure of the potential landscapes.

A fair survey of the extrapolation of the models to large mass 
number and large deformation has therefore to cover a selection 
of typical parameterizations.  We have chosen parameterizations 
which give a very satisfactory description of stable nuclei but 
differ in details. Namely we use the Skyrme interactions SkP \cite{SkP}, 
SLy6 \cite{SLyx}, SkI3, and SkI4 \cite{SkIx}. In the relativistic
calculations the parameterizations NL3 \cite{NL3} and NL-Z2 \cite{Ben99a} 
of the standard Lagrangian are employed. 

The parameterization SkP has the isoscalar effective mass 
\mbox{$m^*_0/m = 1$} and was originally 
designed to describe the particle-hole and particle-particle channel 
of the effective interaction simultaneously (we do not make use of this 
particular feature). The forces SLy6, SkI3, and SkI4 stem from recent 
fits including already data on exotic nuclei (all three forces) and
even neutron matter (SLy6). Both SkP and SLy6 use the standard spin-orbit 
interaction. The forces SkI3/4 employ a spin-orbit force with modified 
isovector dependence. SkI3 contains a fixed isovector part analogous to 
the non-relativistic limit of the RMF, whereas SkI4 is adjusted allowing
free variation of the isovector spin-orbit term. The RMF force NL-Z2
is fitted in the same way as SkI3 and SkI4 to a similar set of
observables.  
%
%=========
%
\begin{table}[t!]
\caption{\label{tab:asurf}
Compilation of bulk properties for the parameterizations 
employed in this study. The upper block shows the volume
parameters incompressibility modulus $K$, effective mass $m_0^*/m$, 
and asymmetry energy coefficient $a_{\rm sym}$. The lower block 
shows the surface energy coefficient $a_{\rm surf}$ obtained 
from semi-infinite nuclear matter calculations.
For the RMF, where the effective mass is momentum-dependent,
$m_0^*/m$ is given at the Fermi momentum \mbox{$k=k_{\rm F}$}. 
This value, which is about 10$\%$ larger than the often quoted 
value for \mbox{$k=0$}, determines the average level density 
around the Fermi energy, see also Ref.\ \cite{Ben99a} and 
references therein.
}
\begin{tabular}{lcccccc}
\hline\noalign{\smallskip}
force                & SkP  & SkI3 & SkI4 & SLy6 & NL-Z2 & NL3 \\
\noalign{\smallskip}\hline\noalign{\smallskip}
$K$ (MeV)            & 202  & 258  & 248  & 230  & 172  & 270 \\
$m_0^*/m$            & 1.00 & 0.59 & 0.65 & 0.69 & 0.64 & 0.67 \\
$a_{\rm sym}$ (MeV)  & 30.0 & 34.8 & 29.5 & 32.0 & 39.0 & 37.4 \\
\noalign{\smallskip}\hline\noalign{\smallskip}
$a_{\rm surf}$ (MeV) & 18.2 & 18.3 & 18.3 & 17.7 & 17.7 & 18.5 \\
\noalign{\smallskip}\hline
\end{tabular}
\end{table}
%
%=========
%

A quantity that characterize the average deformation properties 
of an effective interaction is the surface energy coefficient. 
It is determined for the model system of semi-infinite nuclear 
matter, which offers the cleanest procedure to define a surface 
energy, see, e.g.\ \cite{RMP,Far82,Eif94} and the references 
given therein.  In table \ref{tab:asurf}, we give values for bulk 
and surface properties of nuclear matter obtained 
exclusively in fully self-consistent Hartree-Fock calculations of
semi-infinite matter \cite{Stopriv,Sampriv}. Often values obtained
within the extended Thomas-Fermi approximation are given \cite{Bra85aR},
which are usually smaller by about 1 MeV.

A correction of the binding energy for spurious center-of-mass motion
is performed as usual. For SkI3, SkI4, SLy6, and NL-Z2 the
c.m.\ correction \mbox{${E_{\rm c.m.}=\langle \hat\textbf{P}{}^2_{\rm
c.m.} \rangle / 2mA}$} is subtracted after variation. For SkP the
diagonal part of $E_{\rm c.m.}$ only is considered before variation,
while for NL3 the harmonic-oscillator estimate for $E_{\rm c.m.}$ is
subtracted, see Ref.\ \cite{Ben00b} for details.  The c.m.\ correction,
however, varies only little with deformation, see e.g.\ Ref.\
\cite{Ben00b,War02a}. Note that the various recipes for c.m.\
correction cannot be easily interchanged as their differences are
partially absorbed into the force parameters. This has a consequence
relevant for our study: The difference between the ``exact'' and the
approximate schemes used for SkP and NL3 scales in leading order as
$\sim A^{2/3}$. During the fit this difference is incorporated into
the effective interaction which leads to the significantly larger
surface tension for SkP and NL3 found in table \ref{tab:asurf}, see Ref.\
\cite{Ben00b}.  The deformation energy from otherwise equally fitted
forces might differ on the order of 5 MeV at the outer barrier in
actinides. For SHE, where the saddle-point is at smaller deformation,
this effect can be expected to be less pronounced, but still might
cause differences of a few MeV between forces.

We treat pairing correlations within the BCS approximation using an
effective density-independent zero-range delta pairing force with the 
strength adjusted for each mean-field parameterization separately as 
described in Ref.\ \cite{Ben00c}. Including a density dependence of 
the effective pairing interaction or an approximate particle-number 
projection might alter the barrier heights on the order of 1 MeV 
for actinide nuclei \cite{BenDiss}. 

The coupled mean-field equations for both SHF and RMF models are
represented on a grid in coordinate space using a Fourier
representation of the derivatives and are solved with the damped
gradient iteration method as described in Ref.\ \cite{dampgrad}.  The
numerical codes for both models \cite{RutzDiss,BenDiss} share the same
basic numerical routines which allows for a direct comparison of the
results. Note that the accuracy of grid techniques is fairly
independent on deformation, which is an advantage to calculations
using an harmonic oscillator basis expansion, see e.g.\ Ref.\ 
\cite{War02a} for a convergence study for $^{256}$Fm.

Finally, beyond-mean-field effects can modify the fission barrier.
The most important corrections to the binding energy remove the
contributions from spurious collective vibrational and rotational
states which are inevitably admixed to the mean-field wave functions
\cite{zpeII,War02a,RMP}. These corrections lower the barriers,
typically up to 1 MeV for the inner one and up to 2 MeV for the outer
when starting from a well-deformed prolate ground state. The
situation is less clear for transitional or spherical nuclei.

Altogether, the present study has uncertainties on the deformation
energy for given configuration of the order 1-2 MeV. Most of the 
possible improvements increase the binding energy. Thus we can assume 
to explore an upper limit for the barriers. The comparison of barriers 
between different forces is more robust because most corrections 
can be expected to be similar for all forces.
%
%---------------------------------------------------------------------------
%
\subsection{Shape Degrees of Freedom}
\label{Subsect:shapes}
We will consider reflection-symmetric as well as reflection-asymmetric 
fission paths. But we restrict the considerations to axially symmetric 
shapes. Our investigation also covers the prolate fission path only. 
It has to be kept in mind that novel fission paths may emerge for the 
heaviest of the nuclides discussed here, which start out from strongly 
oblate shapes and proceed through triaxial deformations \cite{earlytriax}.
For this and the reasons given above our results provide an
upper limit for the (static) fission barriers. This limitation holds 
also for most other work using self-consistent models published so far, 
as well as most of the results from mic-mac approaches.

The deformation energy curves are obtained with a constraint on the mass 
quadrupole moment $Q_{20} = \langle \hat{Q}_{20} \rangle$. For 
reflection-asymmetric shapes, we also fix the center-of-mass 
with a constraint on the mass dipole moment 
\mbox{$\langle \hat{Q}_{10} \rangle = 0$}. The constraints
are added to the energy functional by means of Lagrange multipliers 
\cite{RMP}. Besides these constraints, the deformation energy 
is minimized with respect to all axial multipole moments $Q_{\ell 0}$ for 
protons and neutrons separately. In a self-consistent calculation, the 
energy is not only minimized with respect to the deformation, but also the
radial profile of the density distribution, again separately for protons 
and neutrons. With that, a self-consistent calculation explores many 
more degrees of freedom than the best microscopic-macroscopic calculation
available so far. Some consequences will be discussed in 
section \ref{subsect:saddlepoint} below.

The deformation energy is shown versus the dimensionless multipole 
deformations of the mass density which are defined as
\begin{equation}
\beta_\ell
= \frac{4 \pi}{3 A r_0^\ell} \; \langle r^\ell \, Y_{\ell0} \rangle
\quad \hbox{with $r_0 = 1.2 \, A^{1/3}$ fm.}
\end{equation}
Note that the $\beta_\ell$ are computed from the expectation values of
the actual shapes and need to be distinguished from the generating
moments which are used in multipole expansions of the nuclear shape in
microscopic-macroscopic models \cite{Has88a}.

The constrained calculation does not always follow exactly the static
fission path, which is defined as the path that follows the steepest 
descent in the multidimensional energy surface. Instead, for each value 
of $Q_{20}$, one obtains a state which corresponds to a minimum with 
respect to all other degrees of freedom \cite{Flo73}.
This might cause some problems to keep track of the path whenever the 
fission path has a small component only in the direction of the constraint.
Often there exist two or even more distinct valleys in the multidimensional 
potential landscape which are separated by potential barriers. Depending 
on the choice for the initial wave functions, the constrained calculation 
might find the nearest relative minimum only, which is not necessarily the 
absolute minimum for a given constraint. The existence of distinct 
valleys complicates the interpretation of the deformation energy curves. 
When the solution jumps from one valley to another it misses the saddle 
point in between. The resulting uncertainty is not clear \emph{a priori}, 
as only a calculation including two or even more constraints can 
clarify if there is a flat plateau or a mountain ridge between the two
valleys. The change from one to another valley in the potential landscape 
is accompanied by discontinuities in higher multipole deformations
which can be used to identify them.
In some cases the existence of two distinct valleys might be the 
artifact of imposed symmetries, c.f.\ the case of $^{258}$Fm,
where the two distinct valleys obtained from axially symmetric
calculations are spurious as the two solutions are smoothly 
connected through triaxial shapes \cite{Ben98a}.
%
%===========================================================================
%
\section{Barriers in Actinide Nuclei}
\label{sect:actinides}
%
%-----------------
%
\begin{figure}[t!]
\epsfig{figure=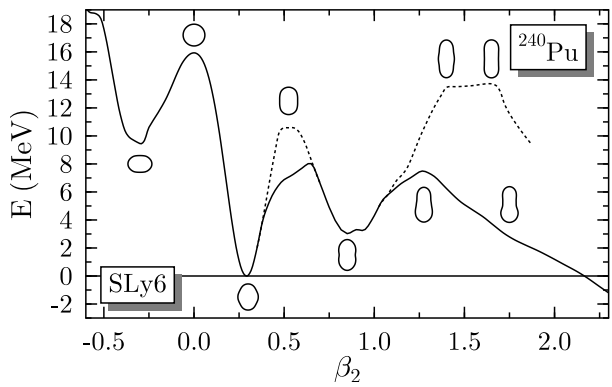}
\caption{\label{pu240}
Example for the double-humped fission barrier of the typical actinide
nucleus \nuc{240}{Pu}. The dotted line denotes an axial and 
reflection-symmetric calculation, the full line denotes a triaxial 
(inner barrier) and axial and reflection-asymmetric calculation 
(outer barrier). The various shapes along the axial paths are indicated
by the contours of the total density at \mbox{$\rho_0 = 0.07$} fm$^{-3}$.
}
\end{figure}
%
%-----------------
%
Actinides are the heaviest nuclear systems for which data on the
structure of the fission barrier are available. We use these nuclei to
benchmark our models and forces, and to examine the force dependence
of the predictions. We confine our investigation to axial barriers but
release reflection symmetry in the calculations of the outer barriers
and isomeric states.

There exists a wealth of information about the (in most cases)
double-humped fission barriers of actinide nuclei, see Ref.\ \cite{Spe74a}
and references given therein. The generic features of the static 
fission path are shown in figure \ref{pu240} for the example of 
\nuc{240}{Pu} calculated with the Skyrme force SLy6. The deformed 
ground-state has a calculated deformation of 
\mbox{$\beta_2 = 0.29$}, which is in perfect agreement with 
the value of \mbox{$\beta_2 = 0.29$} that can be deduced within 
the rigid rotor model from the $B(E2)\uparrow$ value of 
$13.33 \pm 0.18$ $e^2$ b$^2$ obtained from Coulomb excitation 
\cite{Bem73aE}. The deformation energy of 15.9 MeV of the 
ground state corresponds to 0.9 $\%$ of the total binding energy. 
The inner barrier explores triaxial degrees of freedom, which 
reduce the (axial) barrier by about 3 MeV. There is a superdeformed 
fission isomer at \mbox{$\beta_2 \approx 0.8$} at an excitation 
energy of 3.0 MeV, which is somewhat larger than the experimental
value of $2.25 \pm 0.20$ MeV \cite{Hun01}. 
The outer barrier explores reflection-asymmetric shapes. The potential
landscapes of adjacent actinide nuclei are similar, though for some 
nuclides there might appear a second isomeric state \cite{Zha86aE}.

The double-humped fission barrier of \nuc{240}{Pu} has often served 
as a benchmark for mean-field models, see Ref.\ \cite{Flo74} for 
results obtained using Skyrme interactions, \cite{Gir83} using
Gogny forces and Refs.\ \cite{Blu94a,Rut95a} for the RMF, see 
also \cite{Ben00b}.
Early comparisons of barriers obtained with different Skyrme forces
using approximations to full self-consistency were published 
in Refs.\ \cite{Dut80a,Bra85aR,Bar82}.
A direct comparison of the potential energy curves with triaxial inner 
and reflection-asymmetric outer barriers obtained with the Skyrme 
interactions SLy6, SkM* and SkI4, the Gogny force D1s and the RMF 
forces NL3 and NL-Z2 is presented in \cite{RMP}. The excitation 
energy of the fission isomer has been studied with a variety 
of Skyrme forces in Refs.\ \cite{Hee97a,Tak98a}.

The influence of correlations on the deformation energy in the 
framework of Skyrme mean-field calculations has been studied in 
Ref.\ \cite{BHpu240}. An exact angular momentum projection lowers 
the axial inner barrier by a bit less than 1 MeV, the excitation 
energy of the isomeric minimum by about 1 MeV, and the reflection-symmetric
outer barrier by about 1.5-2 MeV. Removing spurious quadrupole 
vibrations from the mean-field states by configuration mixing of
the angular-momentum projected mean field states lowers both the
ground state and the isomeric state by a few 100 keV. For $^{240}$Pu, 
this effect is more pronounced for the isomer, which lowers its
excitation energy even further, but it cannot be expected that 
this will be the same for all actinides.

In this section we will confine ourselves to the heights of the inner
and outer barrier as well as the excitation energies of the isomeric
states while postponing a thorough discussion of the potential
landscapes to future work. As our goal is the extrapolation of the
models to superheavy nuclei, a discussion of these key quantities
suffices.

The selection of even-even nuclei for this study is
$^{230}_{90}$Th$_{140}$,
$^{234}_{90}$Th$_{144}$,
$^{234}_{92}$U$_{142}$,
$^{238}_{92}$U$_{146}$,
$^{238}_{94}$Pu$_{144}$,
$^{242}_{94}$Pu$_{148}$,
$^{246}_{94}$Pu$_{152}$,
$^{242}_{96}$Cm$_{146}$,
$^{246}_{96}$Cm$_{150}$,
$^{250}_{96}$Cm$_{154}$,
$^{250}_{98}$Cf$_{152}$,
which is only every second known nucleus in neutron number.
We omit Fm isotopes as there are two competing paths at the 
outer barrier, and there is no continuous axial path for the
inner barrier for some Skyrme forces \cite{Ben98a}. 
%
%-----------------
%
\begin{figure}[t!]
\epsfig{figure=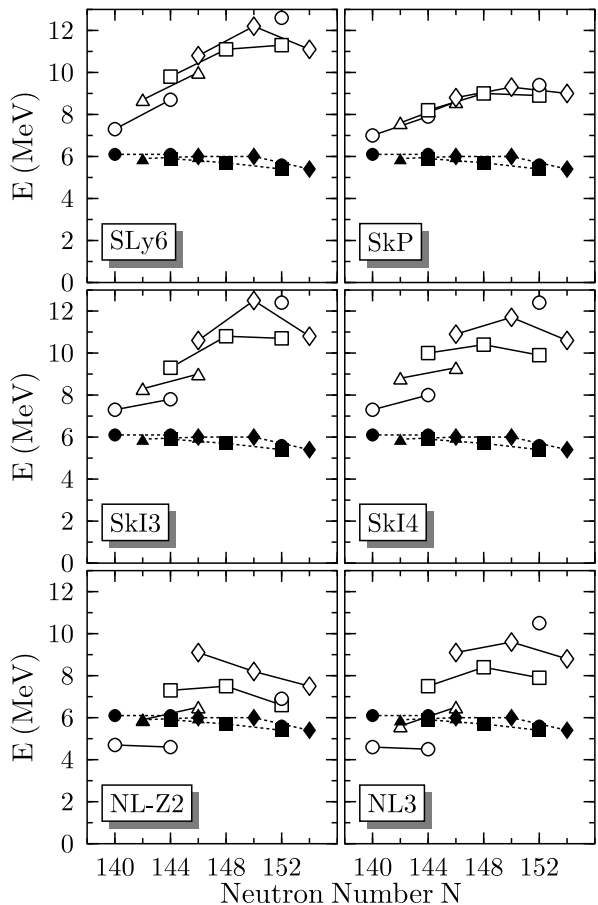}
\caption{\label{1stbar}
Height of the inner barrier from axial and reflection-symmetric 
calculations. Th (\mbox{$Z=90$}), U (\mbox{$Z=92$}), Pu (\mbox{$Z=94$}), 
Cu (\mbox{$Z=96$}), and Cf (\mbox{$Z=98$}) isotopes are denoted by 
open circles (for \mbox{$N=140$}, 142), open triangles, open squares, 
open diamonds, and again open circles (\mbox{$N=152$}), respectively. 
Experimental data (full symbols) are taken from \protect\cite{Mam98a}.
Data points for the same element are connected by lines.
}
\end{figure}
%
%-----------------
%
%
%-----------------
%
\begin{figure}[t!]
\epsfig{figure=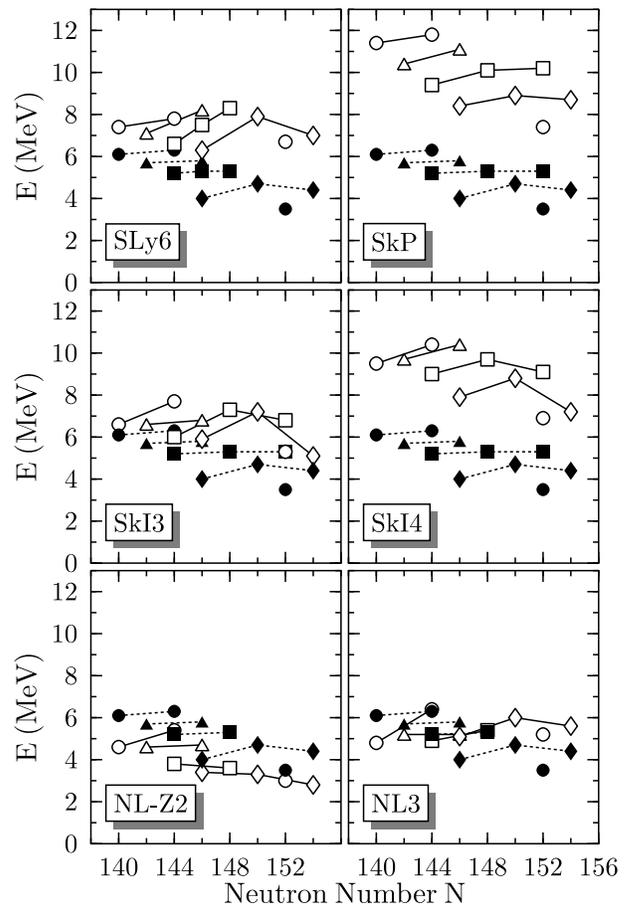}
\caption{\label{2ndbar}
The same as Figure \ref{1stbar}, but for the outer barrier.
}
\end{figure}
%
%-----------------
%

Figure \ref{1stbar} compares calculated and experimental heights 
of the inner barrier. In the case of the inner barrier, the comparison 
of data with values obtained from axial calculations is somewhat 
dangerous, as the static inner barrier is known to be triaxial and 
the energy gain through triaxial deformation may differ for each force.
Still, there are several conclusions that can be safely drawn from 
Fig.\ \ref{1stbar}. Our selection of forces suggests that there is
a difference between SHF and RMF models. All Skyrme forces predict
that the inner barrier increases with neutron number up to 
\mbox{$N=150$}, most pronounced for SLy6. The $Z$ dependence of 
the barrier height is most pronounced for SkI3 and SkI4, the
Skyrme forces with an extended spin-orbit interaction, while it
is negligible for SkP, the only force in our sample with the large 
effective mass \mbox{$m_0^*/m = 1.0$}, c.f.\ table \ref{tab:asurf}.
One might suspect that the corresponding large level density 
suppresses shell effects compared to the other forces. 
On the other hand, for the two RMF forces the barriers stay nearly 
constant with $N$, and show an increase with $Z$ only. This finding
suggests a significant difference in shell structure between the
SHF and RMF models.  Experimental data do not show any significant 
dependence on $N$ or $Z$ at all, they just fall off a few 100 keV 
with mass number. From our present calculations, it cannot be 
decided if adding triaxiality will give a similar trend.
%
%-----------------
%
\begin{figure}[t!]
\epsfig{figure=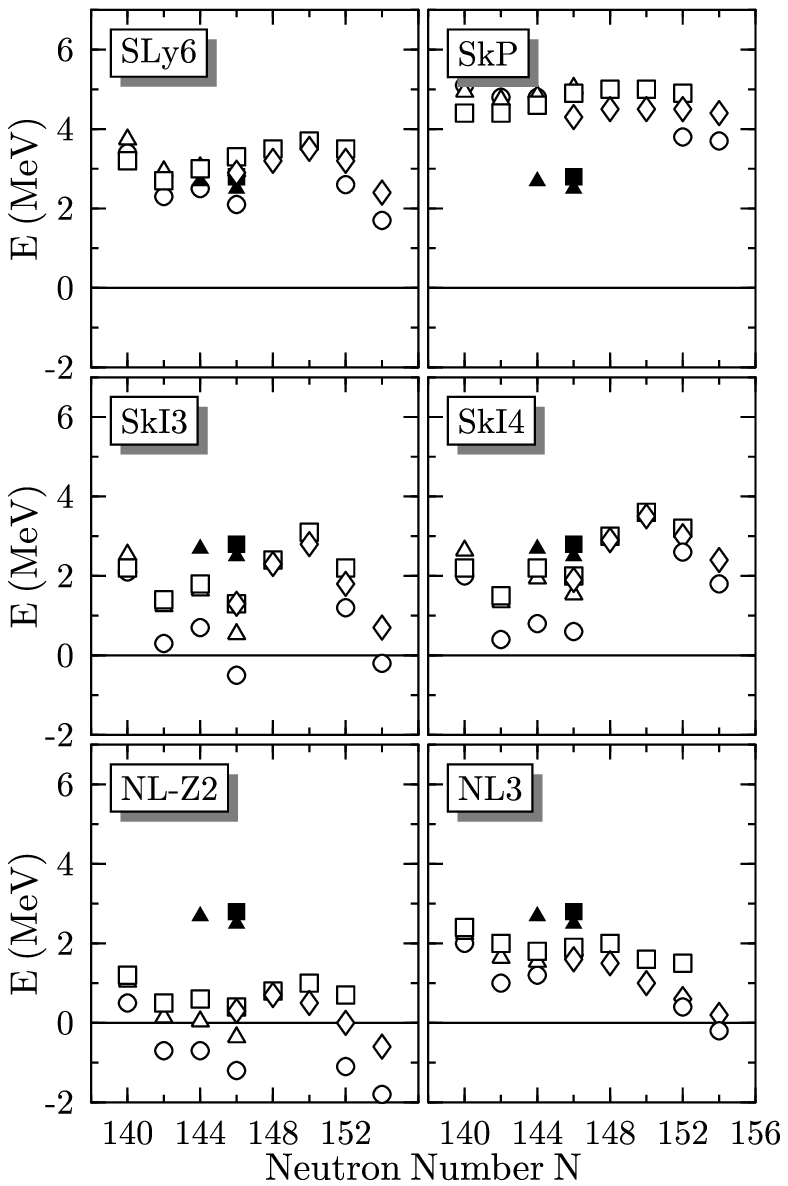}
\caption{\label{act:isomer}
Excitation energy of the isomer, obtained from axial 
calculations.
}
\end{figure}
%
%-----------------
%

There is also a difference in absolute height between SHF and RMF.
With the exception of SkP, the inner barriers from SHF are 
significantly higher by about 2 MeV than those obtained within 
the RMF. This is reflected in the values for the mean deviation of the 
inner barrier height $(1/n) \sum_{i=1}^{n}|\Delta E_i|$ from the 
experimental value in MeV is 4.5 (SLy6), 2.7 (SkP), 4.1 (SkI3), 
4.1 (SkI4), 1.5 (NL-Z2), and 2.3 (NL3) respectively, although this 
quantity is of limited significance. Angular momentum projection
will lower the barrier by about 1 MeV, and one can speculate only
about the effect of triaxiality and other correlations on the trends
with $N$ and $Z$. It is tempting to assume that, when including
these missing corrections, the SHF might still overestimate the 
barriers of the heavier nuclei but be on the right order for the
lighter ones, while the RMF will underestimate the barriers of Th 
isotopes already on the mean-field level.

The outer barrier heights as predicted by the various mean-field
forces are shown in Fig.\ \ref{2ndbar}. For the outer barrier, it 
can be expected that our calculations cover all necessary degrees 
of freedom, so data and calculated values can be directly compared.
There are differences in absolute barrier height. With a mean 
deviation of the outer barrier height in MeV of 
2.2 (SLy6), 4.7 (SkP), 1.4 (SkI3), 3.8 (SkI4), 1.2 (NL-Z2), and
0.8 (NL3) respectively, SkI3, NL-Z2 and in particular NL3 give
a quite good description of the barrier height on the mean-field
level. The differences found for the barrier height seem to manifest 
themselves mainly in an overall offset, while all models and forces predict 
quite similar trends of the barrier heights with $N$ and $Z$, a bit more
pronounced for Skyrme forces, and a bit more damped for NL-Z2. The good
news is that this overall trend is quite close to the experimental 
findings.

As barriers from NL-Z2 are already always smaller than the 
experimental values, there is no room left for correlation effects.
For shapes around the outer barrier, the (missing) rotational correction 
can be expected to be about 1.5 MeV larger than for the ground state 
of a well-deformed nucleus. When removing those 1.5 MeV from the outer 
barrier heights shown in Fig.\ \ref{2ndbar}, the barriers from the 
RMF are too low, in particular for NL-Z2, where not much will be left.

The excitation energy of the fission isomer is displayed in
Fig.\ \ref{act:isomer}. We have added also results obtained 
for the nuclei with neutron numbers between those used to 
investigate the barriers above.
To the best of our knowledge, only three experimental values 
for superdeformed $0^+$ states in even-even nuclei are 
available so far from spectroscopy in the superdeformed 
and normal-deformed wells, which are \nuc{236,8}{U} and 
\nuc{240}{Pu}, see the recent collection of data in \cite{isomers}.
There are more superdeformed levels known in some other 
adjacent even and odd nuclei, but their quantum numbers
could not be established so far. They all have in common
that their excitation energy is at least 2 MeV, which 
also sets some constraints to our calculations. These values 
are also consistent with the data obtained from fits to
fission-isomer excitation functions, see, e.g., Ref.\ \cite{Bri71a}.

Although the sparse data do not allow for a detailed analysis, there
are a few conclusions that can be drawn. On the mean-field level, 
SkP gives rather high energies and overestimates the data by about
2 MeV. Values from SLy6 are scattered around the data. RMF models 
predict very low excitation energies. The Skyrme interactions 
SkI3 and SkI4 with extended spin-orbit interactions also underestimate 
the known excitation energies at least for certain elements.
It is noteworthy that SkI3, NL3 and particularly NL-Z2 
predict the superdeformed state to be the ground state for some 
actinide nuclei, in contradiction with experimental knowledge about 
the spectroscopy and decay of those nuclei. It was already noticed in 
Ref.\ \cite{Rut95a} for selected examples that the RMF underestimates 
the excitation energy of the fission isomer. This seems to be
a general shortcoming of the RMF model, at least of most, if not 
all, of its standard parameterizations. This finding is not restricted 
to actinide nuclei, but was also observed for superdeformed states
in the neutron-deficient \mbox{$A \approx 190$} region
\cite{Hey96a,Nik02a}, where it can be cured to some extend taking
additional information about the spherical shell structure of
\nuc{208}{Pb} into account during the fit of the force parameters
\cite{Nik02a}. The overbinding of the fission isomer can be expected 
to become even more pronounced when corrections for breaking 
of rotational and other symmetries are considered.

From our small selection of Skyrme forces, it is hard to 
disentangle the influence of the effective mass and of the 
various spin-orbit functionals from the influence of the 
actual fitting procedure on the predictions for barrier heights.
It is tempting to correlate the large deformation energy
obtained with SkP with its large effective mass, but this is ruled
out by recent Skyrme-HFB mass fits with effective mass around 1.0 
that deliver much smaller barriers \cite{samyn} than SkP.
Since NL-Z2 and NL3 have the same functional form, the tendency
of NL3 to larger barriers must result from the different strategies 
to adjust the force. 

It is a bit surprising that the differences in surface tension
visible in table \ref{tab:asurf} are not reflected in the inner
barrier heights and cannot be solely responsible for the differences
obtained for outer barrier heights and fission isomers. Shell
effects seem to play a much larger role for these quantities than
the nuclear matter properties. For the forces in our sample, the 
difference in surface tension seems to be compensated by other features 
of the forces through the fit. However, when the influence of
shell effects is suppressed by comparing predictions of pairs of 
otherwise identically fitted forces with different surface
tension, one finds indeed the expected difference, see Ref.\ 
\cite{Ben00b} for the example of the Skyrme interactions 
SLy4 and SLy6 and the RMF forces NL1 and NL-Z.

Summarizing, a differences in the models, can be recognized: 
SHF gives usually higher barriers than RMF. In comparison 
with experimental data, it seems that the RMF predictions 
are too low for barriers and fission isomers. These findings 
were already hinted in earlier investigations of actinide 
\cite{Rut95a,Ben00b} and superheavy nuclei \cite{Ben98a}, 
but emerge even more clearly for the present systematic 
investigation of nuclei.

In the next section we will see how these trends translate to 
superheavy nuclei.
%
%===========================================================================
%
\section{Potential Landscapes}
\label{sec:lands}
%
%=============
%
\begin{figure*}[t!]
\epsfig{figure=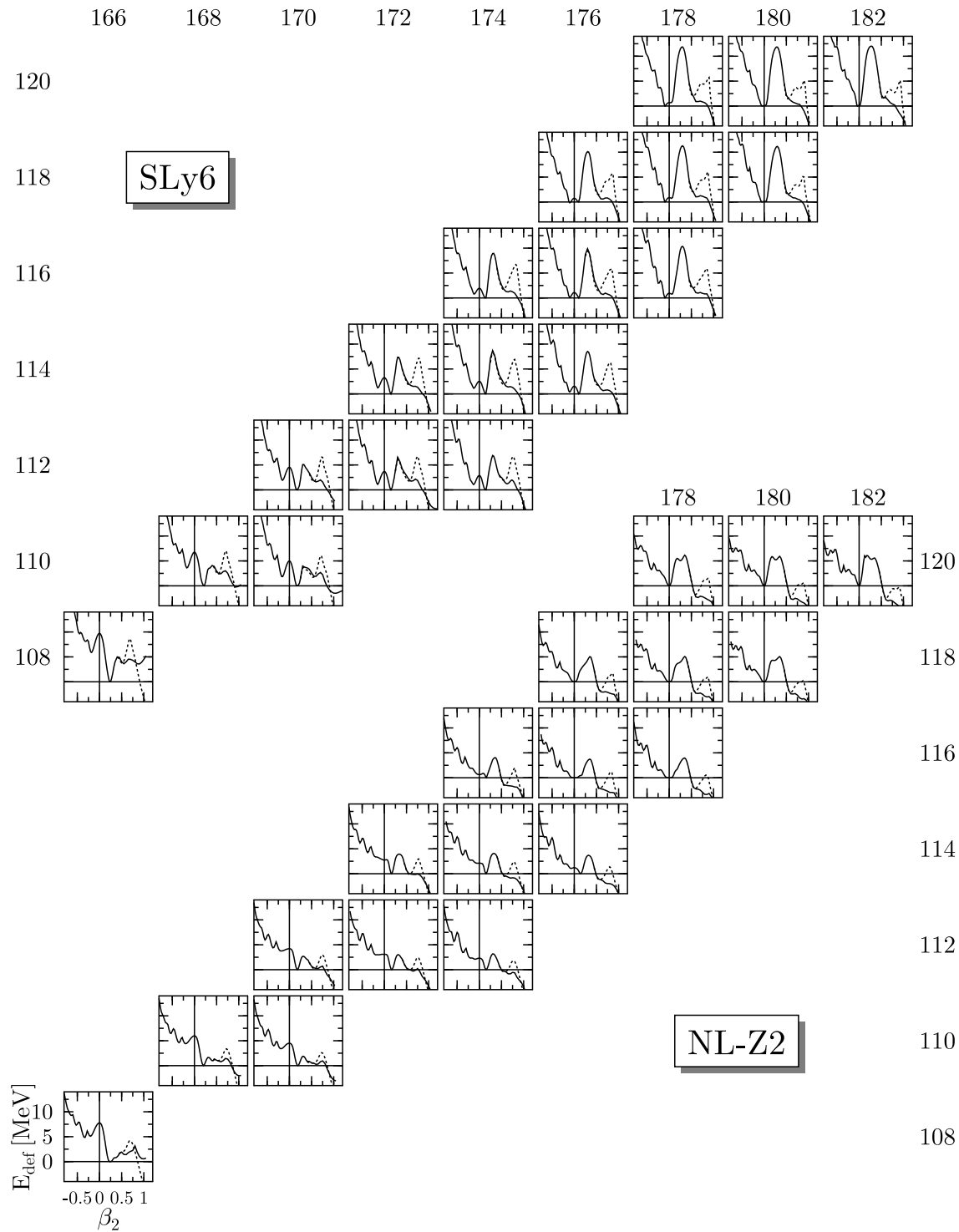}
\caption{\label{fig:all-sly6-nlz2}
Axial fission barriers for the Skyrme force SLy6 (top) and the 
relativistic force NL-Z2 (bottom). Solid (dashed) lines denote 
the reflection-asymmetric (reflection-symmetric) path.
}
\end{figure*}
%
%=============
%
Figure \ref{fig:all-sly6-nlz2} provides a summary
view of the deformation energy curves along the fission paths for
all SHE under consideration here for the Skyrme interaction SLy6 and 
the relativistic mean-field force NL-Z2. The full lines denote the 
asymmetric fission path, the dashed lines the symmetric fission path
(which coincide at small deformations).
The global trends are common for both forces (and also the others 
employed in this study) and in qualitative agreement with earlier
studies in mic-mac and semiclassical models: 
\begin{itemize}
\item There is a gradual transition from well-deformed nuclei 
      with \mbox{$\beta_2 \approx 0.3$} around the deformed 
      \mbox{$Z=108$} and \mbox{$N=162$} shell closures to 
      spherical shapes approaching \mbox{$N=184$}. 
      Note that earlier studies suggest that the neutron number 
      is more important than the proton number to determine the 
      ground-state shape.
\item Intermediate systems around \mbox{$Z=114$}, \mbox{$N=174$} have
      two distinct prolate and oblate minima at small deformation. On
      the basis of our axial calculations it is not clear if this 
      leads to shape coexistence in these nuclei as concluded on 
      similar grounds in Refs.\ \cite{Pat99a,Ren01a,Ren02a}. 
      Calculations including triaxial degrees of freedom suggest that for
      some, perhaps all of these systems the prolate and oblate ``minima''
      are connected through triaxial shapes without a barrier \cite{Cwipc}.
\item While the ground-state deformation moves to smaller values, also 
      the saddle point is shifted to smaller deformations. This means that
      the width of the fission barrier is not necessarily larger for
      nuclei with spherical shell closures. 
\item The transitional nuclei in between belong to a regime of low fission
      barriers. This is consistent with the current (still sparse) 
      experimental knowledge. The recent data from Dubna interpreted as the 
      $\alpha$-decay chain of $^{292}$116 indicate that fission is 
      the preferred decay channel of $^{280}$110$_{170}$, but not the 
      heavier nuclides in this chain \cite{exp116}.  
\item Above \mbox{$Z=108$}, the static fission path switches from 
      symmetric to asymmetric fission. 
\item For most of the nuclei above \mbox{$Z \geq 110$} considered here, 
      reflection-asymmetric shape degrees of freedom remove completely 
      the outer barrier that is well-known from actinide nuclei, 
      leading to a single-humped fission barrier only.
\item The superdeformed minima around \mbox{$\beta_2 \approx 0.5$}
      obtained from reflection-symmetric calculations with NL-Z2 
      for $^{298}118_{176}$ in Ref.\ \cite{Ren01a} and $^{292}116_{176}$, 
      $^{288}114_{174}$, $^{284}112_{172}$, and $^{280}110_{172}$ 
      in Ref.\ \cite{Ren02a} are not stable with respect to octupole 
      distortions, which makes the conclusions about superdeformed 
      ground states of superheavy  nuclei drawn in Ref.\ \cite{Ren01a,Ren02a}
      questionable. This finding is not completely general as for some
      forces there remains a very small asymmetric barrier.
\end{itemize}
The general features of these trends are easily understood in the more
intuitive language of the mic-mac models (although they apply, of
course, to the self-consistent models as well). For nuclei at the lower 
end of the region investigated here, the potential energy surface
from the LDM is rather flat around the spherical point and drops off 
fast at prolate deformations
about \mbox{$\beta_2 \approx 0.7$}, c.f.\ figure~7 in Ref.\ \cite{Cwi96a}.
The structures seen in Figures \ref{fig:all-sly6-nlz2}
are mainly determined by the variation of the shell correction with 
deformation. The maximum of the shell correction follows the shell
closures from the deformed \mbox{$Z=108$} and \mbox{$N=162$} to the
spherical \mbox{$N=184$} shell. The potential wells are deepest 
in the vicinity of closed shells. 
With increasing $Z$, the plateau is shifted toward oblate deformations, 
while the LDM surface drops at smaller and smaller prolate deformations, 
which cannot be counterweighted by the variation of $E_{\text{shell}}$. 
With that the saddle point moves in toward smaller deformations.
For larger systems than those discussed here the potential 
energy surface becomes also unstable on the oblate side.

Figure \ref{fig:all-sly6-nlz2} gives also an idea where triaxiality 
might play a significant role: Whenever the deformation energy is 
smaller on the oblate side than for the same deformation on the 
prolate side outside the prolate saddle point, oblate shapes might 
be unstable through a triaxial path. Of course this is neither a 
sufficient nor necessary condition.
%
%=============
%
\begin{figure}[t!]
\epsfig{figure=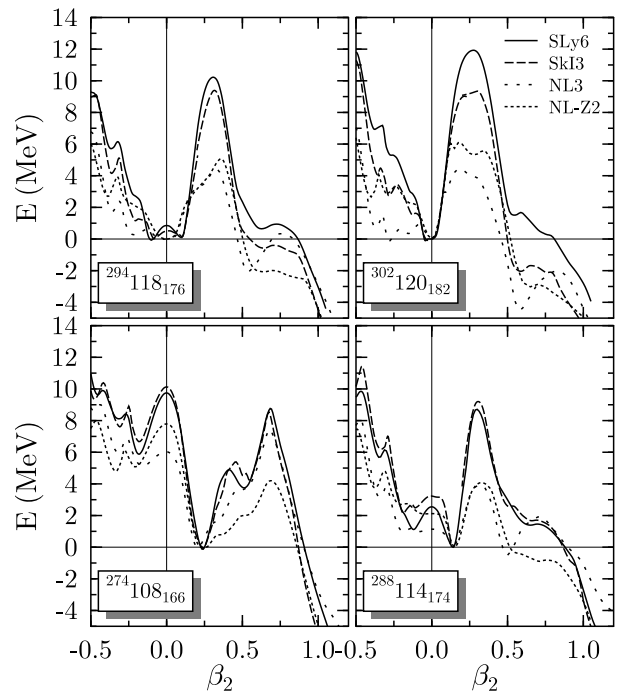}
\caption{\label{barriers-4forces}
The fission barriers of four selected nuclei normalized to the
ground-state for four mean-field parameterizations as indicated.
Shown is always the energetically favoured fission
barrier.}
\end{figure}
%
%=============
%

While there is overall qualitative agreement among the two forces 
(and models), there are significant differences at a quantitative
level. The RMF force NL-Z2 predicts lower barriers when going 
toward heavy systems than the Skyrme interaction SLy6. 
This is confirmed when directly comparing the deformation energy for 
selected nuclei, see Fig.\ \ref{barriers-4forces}. 
$^{274}_{166}$Hs$_{108}$ is a well-deformed nucleus located at the 
edge of the ``rock of stability'' around  $^{270}_{162}$Hs$_{108}$,
$^{302}_{182}$120 is close to the spherical neutron shell \mbox{$N=184$},
while the other two are located in the transitional region.
The figure compares more forces, now two from SHF (SLy6 and SkI3) 
and two from RMF (NL3 and NL-Z2). 

There are two kinds of differences: first, the systematic difference 
between SHF and RMF models which we saw already for the actinides, 
with the RMF giving smaller barriers, persists to superheavy systems, 
and second, an additional difference between the two RMF parameterizations 
NL-Z2 and NL3 concerning the outer barrier occurs, 
with NL3 being the only force predicting a double-humped barrier for
the heavier systems.

Comparing the potential energy curves from NL-Z2 and SLy6 at small
deformation for the heaviest nuclei, one sees also some differences 
concerning how strongly the nuclei are driven to sphericity. Comparing
nuclides in the ``northeastern'' corner of figure 
\ref{fig:all-sly6-nlz2}, the onset of spherical ground states is
predicted to be earlier with NL-Z2 than with SLy6.
This reflects the different predictions for shell closures from 
both models \cite{Rut97a,Ben99a}. While NL-Z2 (like all other 
standard RMF forces) predicts strong \mbox{$Z=120$} and \mbox{$N=172$}
shells and a weak \mbox{$N=184$} shell, SLy6 gives a strong 
\mbox{$N=184$} shell and a weak \mbox{$Z=120$} shell, which 
is not sufficient to guarantee a spherical ground state of 
this nucleus for non-magic neutron number. Therefore nuclei at the
upper end of figure \ref{fig:all-sly6-nlz2} are much more driven 
to sphericity when calculated with NL-Z2 than with SLy6.
%
%============================================================================
%
\section{Characteristic Quantities}
\label{sec:chara}
The basic features of all the deformation energy surfaces shown in the
previous plots can be characterized by a few key numbers, i.e.\
the ground-state deformation energy, the height of the inner fission
barrier and the outer barrier, which will be discussed in this section. 
%
%---------------------------------------------------------------------------
%
\subsection{Ground-State Deformation Energy}
%
%
%-----------------
%
\begin{figure}[t!]
\epsfig{figure=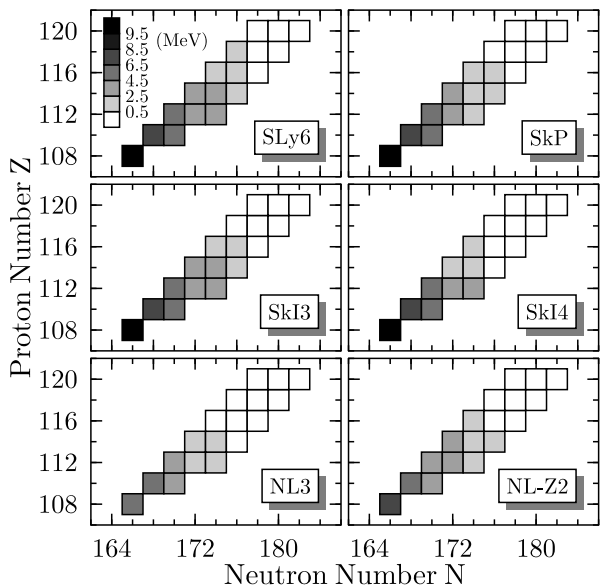}
\caption{\label{defenergy}
Deformation energy for the lowest minimum at small deformation
(\mbox{$\beta_2 \leq 0.4$}) for the nuclides and forces as indicated. 
White squares denote deformation energies smaller than 0.5 MeV. See
the main text for a detailed description.}
\end{figure}
%
%-----------------
%
The ground-state deformation energy, i.e.\ the energy difference 
between the spherical shape and the (possibly deformed) ground state
is plotted in figure \ref{defenergy}. We consider only small 
deformations \mbox{$\beta_2 \leq 0.4$}. Zero deformation energy 
always indicates a spherical ground state, as we do not find 
coexisting well-deformed minima at the same energy as the spherical 
configuration.

We see that all models and parameterizations predict a transition 
from deformed nuclei around \mbox{$Z=108$}, \mbox{$N=162$} to 
spherical \mbox{$N=184$} in agreement with earlier studies
\cite{Cwi96a,Bue98a}. These deformation energies correspond to 
strongly prolate deformations at the lower corner of our selection 
of nuclei. For higher $Z$ values, a shape isomerism is established 
with two minima on the prolate and oblate side having approximately 
the same energy and only small deformations of $\beta_2 \approx \pm 0.15$.
Thus far the general trends agree. There are differences in
quantitative detail, most prominently the fact that RMF predicts
systematically lower deformation energies than SHF.

It is to be noted that in SHE the neutron number most often determines
the ground-state shape, while a magic proton number might not prevent
deformation. This seems to be a general feature of all self-consistent 
models \cite{Cwi96a,Bue98a}, and can also be observed in mic-mac models,
which predict deformed \mbox{$Z=114$} isotopes \cite{Mol94a,Mol97a}
far off \mbox{$N=184$}. This is due partially
to the overall larger shell correction energy of the spherical
neutron shells compared to the proton shells \cite{Kru00a}, and
partially to the existence of many deformed proton shell
closures in the region \mbox{$108 \leq Z \leq 120$}, which drive
nuclei with non-magic neutron number toward deformation.
%
%---------------------------------------------------------------------------
%
\subsection{Existence of Shape Isomers}
%
%------------
%
\begin{figure}[t!]
\epsfig{figure=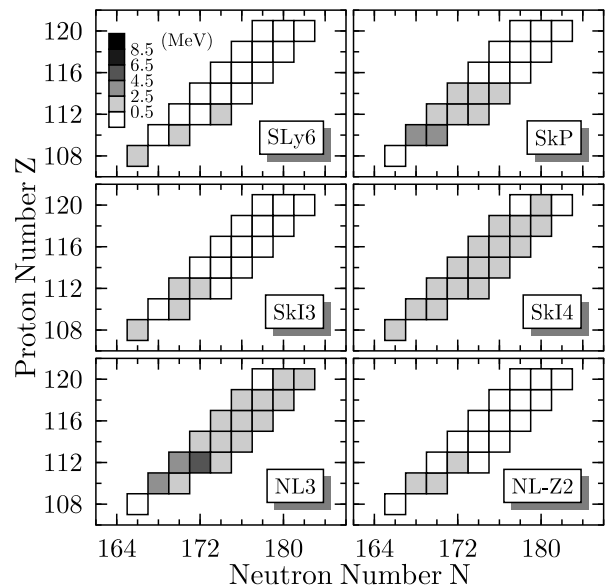}
\caption{Height of the outer, usually asymmetric, fission barrier
with respect to the isomeric state. White squares indicate a 
second bump smaller than 0.5 MeV.}
\label{isomer} 
\end{figure}
%
%-----------
%
The existence of a shape (or fission) isomer with deformation around
\mbox{$\beta \approx 1.0$} is a prominent feature of actinide nuclei 
\cite{Spe74a}. A fission isomer necessarily requires an outer 
fission barrier. The height of the outer barrier with respect to 
the isomeric state is shown in Fig.\ \ref{isomer}. All forces and 
models confirm the earlier finding that the outer barrier fades 
away for transactinide nuclei, see e.g.\ Ref.\ \cite{Ben98a}. 
An exception is NL3, which predicts a substantial outer barrier
for most nuclei, c.f.\ also figure \ref{barriers-4forces}. As NL-Z2
usually does not show a fission isomer, this cannot be a general 
feature of the RMF model, but has to be a particularity of the NL3 
parameterization. Similarly, SkI4 is an exception among the Skyrme
interactions. Most nuclear matter properties of NL3 and NL-Z2 are
very close, the same holds for the Skyrme interactions SkI4, SkI3 
and SLy6. This suggests that the height of the outer barrier is
mainly determined by shell structure, not the average liquid drop
properties. Remember that SkI4 employs a non-standard spin-orbit 
interaction which leads to single-particle spectra different from those 
of the other self-consistent models at spherical shape \cite{Ben99a}.
%
%---------------------------------------------------------------------------
%
\subsection{Saddle-Point Height}
\label{subsect:saddlepoint}
%
%=============
%
\begin{figure}[t!]
\epsfig{figure=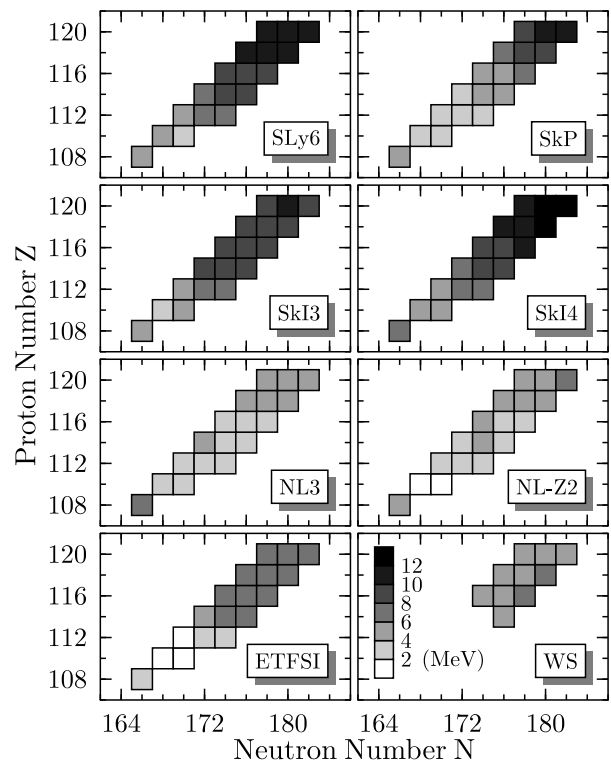}
\caption{The height of the symmetric inner barrier calculated in axial 
symmetry for the models and forces as indicated. We have added also 
ETFSI taken from Ref.\ \protect\cite{Mam01a} and mic-mac results obtained
within the YPE+WS model from Ref.\ \protect\cite{Smo97a}.
}
\label{barriers}
\end{figure}
%
%=============
%
A most interesting feature of SHE is the stability against spontaneous
fission. The fission half-live can be computed from a tunneling
dynamics in the shape degrees of freedom \cite{Nil95a}. Input to that are the
collective masses along the fission path and the fission barriers. We
aim here at a mere comparison of stability between the different
forces. To that end, we assume that the collective masses are about
similar in all cases and confine the discussion to the height of the
(inner) axial fission barrier as key quantity. Keep in mind that the numbers
given for the barrier height represent an upper limit due to possible
lowering through triaxial shapes.

The systematics of the fission barriers for all nuclei and mean-field
forces considered here are shown in figure \ref{barriers}. We have
added results from two other large-scale calculations, one employing
the semi-classical ETFSI approach \cite{Mam01a}, and the other within
the mic-mac approach \cite{Smo97a}. In both cases axial shapes which
allow for reflection-asymmetry are assumed, similar as in our
calculations. The mic-mac fission barriers from Ref.\ \cite{Smo97a} are 
dynamical barriers, i.e.\ the barrier which lies on the fission path 
that minimizes the multi-dimensional action.

All models and forces agree that there is a regime of low fission 
barriers around \mbox{$Z=110$}, but the (axial) barriers increase again
when going toward \mbox{$N=184$}. There are, however, significant 
differences among the models. Comparing the fully self-consistent 
models, the barriers from RMF are much lower than those from SHF. 

And even among the various SHF forces, we see
differences in the barriers. It is again SkI4 with its particular
spin-orbit force which shows the largest barriers. Here it is
noteworthy that SkI4 predicts largest stability for \mbox{$Z=120$}
although it places the magic shell closure at \mbox{$Z=114$}
\cite{Ben99a}. This shows once again that ``magicity'' is something
different from stability as was argued also in Ref.\ \cite{Kru00a}. 

The ETFSI calculations give barriers that have about the size
as the ones from the RMF forces, with the difference that they produce
a bit smaller barriers at the lower end and larger regions with more
stable nuclei.  Similarly, the barriers from the mic-mac
rather correspond to the estimates from the RMF forces. The
self-consistent SHF results deliver the highest barriers throughout.

\begin{figure}[t!]
\epsfig{figure=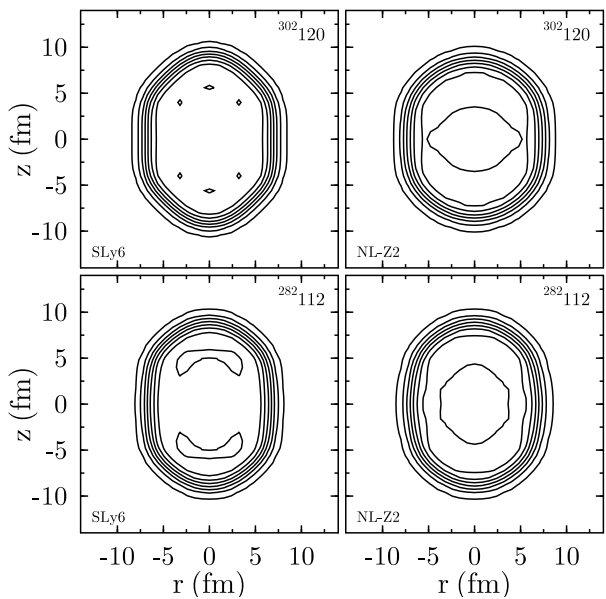}
\caption{\label{fig:saddles}
Contour plots of the mass density distribution for the 
saddle point configurations of the nuclei 
$^{282}$112$_{170}$ (lower panels) and $^{302}120$ (upper panels) 
as predicted by the forces SLy6 (left) and NL-Z2 (right). The
z axis is the symmetry axis. The contour lines correspond 
to the densities 0.01, 0.03, \ldots, 0.15 and 0.17 fm$^{-3}$.
}
\end{figure}

For two nuclei and the forces SLy6 and NL-Z2, the saddle point shapes 
are shown in Fig.\ \ref{fig:saddles}. In all cases, compact nuclear 
shapes are obtained. The saddle point deformation decreases slightly
from values around \mbox{$\beta_2 = 0.32$} obtained with both forces
for $^{274}$Hs, to \mbox{$\beta_2 = 0.28$} obtained with SLy6 and 
\mbox{$\beta_2 = 0.18$} predicted by NL-Z2 respectively for $^{302}120$. 
Note that for NL-Z2 there exists no pronounced saddle point, since the 
barrier is rather flat (see Fig.\ \ref{barriers-4forces}).
%
%=========
%
%
%=========
%

Figure \ref{barriers} also shows fission barriers calculated within
the mic-mac method \cite{Smo97a}. The trends are remarkably different
from the self-consistent models.  A maximum of stability is found
around \mbox{$Z=116$} and less stability for larger systems. Effective mass 
\mbox{$m^*/m=1$} cannot be the reason because SkP with $m^*/m=1$ behaves 
as all other SHF and RMF. We can only speculate about possible explanations.
There might be differences in the smooth part of the self-consistent models,
e.g.\ in higher-order terms missing in the mic-mac models or even the 
curvature term (all of which do not necessarily have the right structure 
in self-consistent models). Or the effect comes from the 
different shell structure between the Folded-Yukawa potential used 
in the mic-mac calculations and self-consistent models \cite{Ben99a} 
(although the difference among, e.g., SkI4 and NL-Z2 is quite
significant, while the global trend of the barriers is not).

It can be speculated that this is a consequence of missing shape
degrees of freedom, either in missing higher-order deformations in the
mic-mac method or in the radial density distribution. It is well-known
that multipole moments at least up to $\ell = 8$ have to be taken into
account to obtain the full shell effect around
$^{270}_{108}$Hs$_{162}$ \cite{Pat91a}. It can be expected that even
more shape degrees of freedom are necessary to get the full shell
effect for the even more complex saddle-point configuration. The
standard mic-mac models also assume that protons and neutrons have the
same deformation. An exploration of the consequences of this
constraint in the framework of the self-consistent Gogny force is
given in Ref.\ \cite{Ber00a}. Imposing the same deformation for
neutrons and protons leads to larger barriers of the order of 1 MeV in
actinides.  A similar effect can be expected for SHE. Even more severe
might be the parameterization of the radial density distribution in
mic-mac models.  There is no radial degree of freedom at all, although
it is well-known that changing the surface diffuseness might
significantly change the ground-state shell correction of SHE. 
The current parameterizations of mic-mac models also prevent
``semi-bubble'' density distributions that might appear at the upper
end of the nuclei in Fig.\ \ref{barriers}.

Although each missing degree of freedom causes a loss in binding energy, 
the deformation dependence of the various effects can be expected to 
be very different. Depending on if the missing energy is larger at 
the ground state or around the saddle point the fission barriers 
are either increased or decreased. This might explain the difference
in the global trend between mic-mac and self-consistent models.

%
%========
%
\begin{table}[t!]
\caption{\label{tab:expt}
Comparison of fission barriers (see text).
Experimental data are taken from Ref.\ \protect\cite{Itk02a}. The
assignment of the neutron number has some uncertainty, therefore
the same experimental barrier appears for two nuclei in the list.
}
\begin{tabular}{lccccccc} 
\hline\noalign{\smallskip} 
nucleus  & Expt. &   NL3  &  NL-Z2 &  SLy6 &  SkI3 &  SkI4 & SkP \\
\noalign{\smallskip}\hline\noalign{\smallskip} 
$^{284}$112$_{172}$ &  5.5  & 3.38  &  2.99  &  6.06 &  6.75 &  6.03 & 2.77 \\
$^{286}$112$_{174}$ &  ''   & 3.41  &  3.16  &  6.91 &  7.52 &  6.97 & 2.77 \\
$^{288}$114$_{174}$ &  6.7  & 3.87  &  4.08  &  8.12 &  8.75 &  8.11 & 4.02 \\
$^{290}$114$_{176}$ &  ``   & 3.56  &  3.70  &  8.52 &  8.15 &  8.67 & 4.31 \\
$^{292}$116$_{176}$ &  6.4  & 3.81  &  3.74  &  9.35 &  8.77 &  9.62 & 5.67 \\
$^{294}$116$_{178}$ &  ``   & 3.80  &  3.96  &  9.59 &  8.61 & 10.93 & 6.50 \\
\noalign{\smallskip}\hline
\end{tabular}
\end{table}
%
%========
%

Table \ref{tab:expt} compares our calculated barrier heights with the lower 
limits of the barriers of some very heavy nuclei recently deduced from 
an analysis of the available data for fusion and fission \cite{Itk02a}. 
Surprisingly these experimentally estimated barrier heights are similar, 
or even slightly larger, than that of actinide nuclei in the $^{240}$Pu 
region. Experimental and calculated values are in agreement for the 
Skyrme forces SLy6, SkI3 and SkI4, while both RMF forces and the Skyrme 
interaction SkP significantly underestimate the barrier.

Although the barrier heights are comparable, the lifetimes corresponding 
to these barriers are much shorter than for actinide nuclei as the 
barriers are much narrower. For the adjacent $^{280}_{170}$110 a 
fission lifetime of about $T_{1/2} = 7.6^{+5.8}_{-2.3}$ s 
was reported in Ref.\ \cite{exp116}. This is quite short but two 
orders of magnitude longer than results of the mic-mac model which 
predict about $10^{-1}$s \cite{Smo97a}. As those heavy nuclei
are solely stabilized by shell effects \cite{Pat89a}, reliable
predictions will be a difficult task for any model.
%
%======================================================================
%
\section{Search for underlying mechanisms}
\label{sec:search}
The above results on the fission barriers and its trends toward the
heaviest SHE show systematic differences between SHF and RMF.  In this
section, we want to ponder a bit about possible reasons.  For the
further discussion it is useful to distinguish between the macroscopic
part of the models (which determines the nuclear matter properties and
the average trends), and the microscopic part (which determines the
actual shell structure). As we will see, it is not yet fully clear
which part is responsible for the observed systematic differences.
%
%----------------------------------------------------------------------
%
\subsection{Macroscopic aspects}
%
%
%=============
%
\begin{figure}[t!]
\epsfig{figure=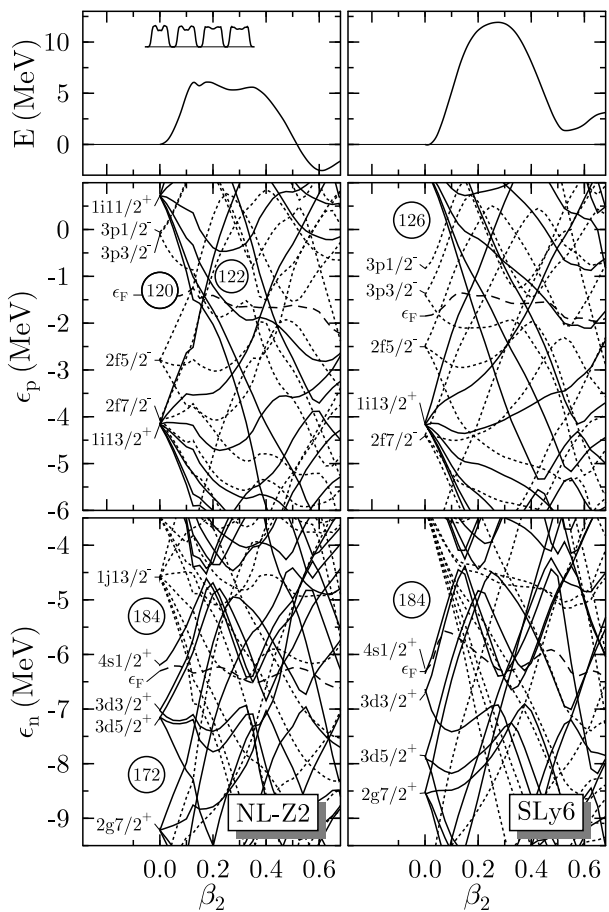}
\caption{\label{levels_182_120} 
Inner axial fission barrier (top), proton levels (middle) and neutron 
levels (bottom) for the nucleus $^{302}120$ with NL-Z2 (left) 
and SLy6 (right). Solid (dotted) lines in the Nilsson plots denote
single-particle states with positive (negative) parity, while the 
dashed line plots the Fermi energy. In the upper panel for NL-Z2
the radial distribution of the total density along the $z$ axis is
also shown.
}
\end{figure}
%
%=============
%
It is well known that most nuclear matter properties from SHF and RMF
models differ significantly, see e.g.\ Ref.\ \cite{RMP} and references
therein. In lowest order, the barrier heights can be expected to scale with
the surface energy coefficient $a_{\rm surf}$. With the exception of 
NL3, however, the values for $a_{\rm surf}$ are quite close for all forces
used here, see Table \ref{tab:asurf}. NL3 gives in most cases 
smaller (inner) barriers in spite of its larger surface energy coefficient.

If the differences seen in Fig.\ \ref{barriers-4forces} are rooted in 
the macroscopic part, another bulk property than $a_{\rm surf}$ has to 
be responsible. It is unlikely that this is the volume energy because 
it is basically independent of the nuclear shape.
The situation is more involved for the volume symmetry energy,
coefficient $a_{\rm sym}$.  At first glance, the volume energy also
scales with the nuclear volume, but there is an implicit surface 
effect due to a correlation between the neutron skin and $a_{\rm sym}$:
the skin increases with increasing $a_{\rm sym}$. A systematic 
study performed in Ref.\ \cite{varyT} suggests that this relation 
is unique. A variation of other isovector properties as the sum rule 
enhancement factor $\kappa_{\rm TRK}$ or the surface-asymmetry 
coefficient $\kappa_{\rm sym}$ leaves the skin unchanged.

And indeed, when we calculate fission barriers with the systematically
varied Skyrme forces from Ref.\ \cite{varyT}, we find that the barriers
increase with decreasing $a_{\rm sym}$, while they do not change
when the isovector effective mass is varied. This is consistent with
our findings for the barriers, where the Skyrme forces which all have
$a_{\rm sym} \approx 32$ MeV have larger barriers than the RMF forces
with $a_{\rm sym} \approx 39$ MeV, c.f.\ table \ref{tab:asurf}.  This
correlation is not unique and might apply only to forces
fitted according to the protocol of Ref.\ \cite{SkyrmeFit}. The Skyrme mass
fit MSk7 \cite{Gor01a} which follows a very different fitting
strategy, has a significantly smaller $a_{\rm sym} = 27.95$ MeV than
the Skyrme forces used here, but predicts also significantly smaller
fission barriers for heavy and superheavy nuclei \cite{samyn}, which
are in fact similar to our RMF results. With that, a satisfying
explanation of the difference between RMF and SHF concerning fission
barrier heights based on macroscopic properties of the
models is still missing.
%
%----------------------------------------------------------------------
%
\subsection{Microscopic aspects}
For nuclei with flat or unstable macroscopic potential energy surfaces
the height of the fission barrier is determined by the variation of the
shell correction energy $E_{\rm shell}$ with deformation. $E_{\rm shell}$
reflects the deviation of the actual density of single-particle levels
around the Fermi energy $\epsilon_{\rm F}$ from an averaged level density.
Pivotal for the barrier is not the absolute value of $E_{\rm shell}$,
but its variation, that reflects the change of the single-particle
spectra with deformation.

Nilsson plots of the single-particle energies of the nucleus 
$^{302}120_{182}$ as calculated with the RMF force NL-Z2 and the Skyrme 
interaction SLy6 are shown in figure \ref{levels_182_120}. There 
are significant differences between the forces which can be traced back 
to their different shell structure at spherical shape (\mbox{$\beta_2 = 0$})
\cite{Rut97a,Ben99a}. 
The small spin-orbit splitting of the $3p$ and $2f$ states obtained with 
NL-Z2 leads to a major shell closure at \mbox{$Z=120$}, while \mbox{$Z=120$}
is a subshell closure only for SLy6, which (for this neutron number)
has more prominent gaps in the single-particle spectrum at \mbox{$Z=114$}
and \mbox{$Z=126$}. On the other hand, there is a huge gap in the
neutron spectrum at \mbox{$N=184$} for SLy6, while there are several
small gaps at \mbox{$N=172$}, 182 and 184 for NL-Z2.

It has to be stressed, however, that for 
self-consistent models such Nilsson plots cannot be extrapolated 
very far from the $N$ and $Z$ they are calculated for. For deformed
shapes, the self-consistent optimization of higher multipole moments 
when changing $N$ and $Z$ might change the single-particle spectra 
significantly. Additionally, the radial shape of the density distribution 
might change with nucleon numbers \cite{Ben99a,Dec99a,Dec03a} or 
deformation. Figure \ref{levels_182_120} provides an example for the 
latter: at small deformation NL-Z2 predicts a semi-bubble shape for 
$^{302}120_{182}$ (see the small inserts in the upper panel), which 
around \mbox{$\beta_2 \approx 0.12$} changes abruptly into a more regular
density distribution, thereby causing the discontinuity in the
single-particle spectra at that deformation. The semi-bubble shape
reduces the spin-orbit splitting (c.f.\ the proton $2f$, $3p$ or
neutron $3d$ states). The appearance of 
such semi-bubble shapes is force dependent, for SLy6 it occurs 
for \mbox{$Z=120$} only at smaller neutron numbers around 
\mbox{$N=172$}. For a more detailed discussion of this phenomenon 
see Refs.\ \cite{Ben99a,Dec99a,Dec03a}.

Comparing the spectra from NL-Z2 and SLy6 at spherical shape,
there is another significant difference besides the spin-orbit
splitting: the highly degenerated $1i_{11/2^+}$ proton and 
$1j_{13/2^-}$ neutron states above the Fermi energy are much lower 
for NL-Z2 than for SLy6. This does not yet lead to a significantly smaller 
(total) shell correction for NL-Z2 (\mbox{$E_{\rm shell} = -13.1$} MeV) 
than for SLy6 (\mbox{$E_{\rm shell} = -14.1$} MeV) \cite{Kru00a},
but their splitting with deformation brings more of these levels close 
to the Fermi energy around the fission barrier for NL-Z2, which might be 
the reason for the difference between NL-Z2 and SLy6. From looking at
the single-particle spectra in figure \ref{levels_182_120} alone, however,
this cannot be decided. To resolve this issue, a calculation of the 
shell correction for deformed shapes along the strategy of 
Ref.\ \cite{Kru00a} seems highly desirable.
%
%=============
%
\begin{figure}[t!]
\epsfig{figure=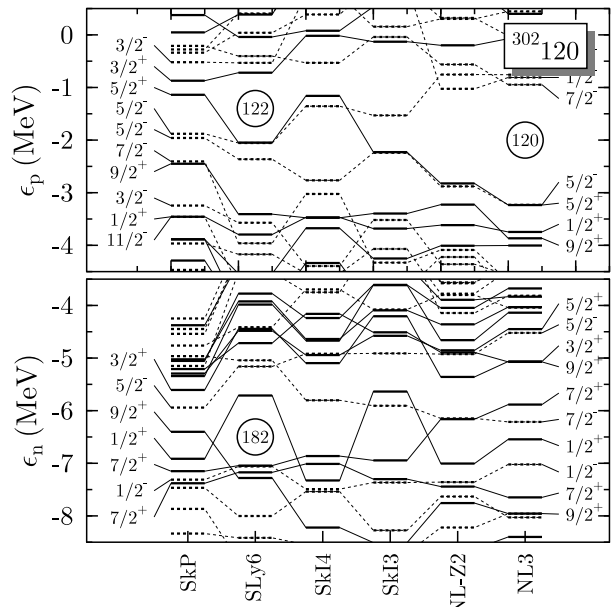}
\caption{\label{182_120:spectra_iso}
Single-particle spectra in the superdeformed configuration obtained 
from reflection-symmetric calculations for $^{302}120_{182}$.
}
\end{figure}
%
%=============
%

The single-particle spectra at the spherical point also determine 
the existence or non-existence of the reflection-asymmetric outer 
barrier, and with that of the fission isomer. At deformations around 
\mbox{$\beta_2 \approx 0.5$}, there are several single-particle states 
originating from the intruder and the major shells above and below 
coming close to the Fermi energy, see Fig.\ \ref{182_120:spectra_iso}
for the example of $^{302}120_{182}$. Spectra for other nuclei in this
region look quite similar. Octupole deformation mixes states with the 
same angular momentum but opposite parity, so depending on the actual 
level ordering it will increase or decrease the level density at 
the Fermi surface. The differences in the relative distance of the 
single-particle states found at spherical shape \cite{Ben99a} are 
reflected in the spectra for these very deformed shapes. For example, 
the subtle shift in single-particle energies between the RMF forces 
NL-Z2 and NL3 removes the outer asymmetric barrier for the former,
but not the latter. The single-particle spectra at spherical shape
including those far above and below the Fermi energy, have to be 
described with very high precision to decide if there 
exist superdeformed states in superheavy elements.
%
%===========================================================================
%
\section{Conclusions}
We have investigated the systematics of fission barriers in superheavy
elements with \mbox{$Z=108$}, \ldots, 120 as predicted by self-consistent 
mean-field models. As typical representatives, we employed the
non-relativistic Skyrme-Hartree-Fock (SHF) model as well as the
relativistic mean-field model (RMF), and for each case, we used a
selection of different parametrizations to explore the variances in
the predictions. All calculations have been done with axial symmetry
but allowing for reflection-asymmetric shapes.

As a benchmark for our mean-field models and forces, a selection of 
actinide nuclei ranging from Th to Cf isotopes has been utilized to 
study the predictions for the inner and outer axial barriers as well 
the excitation energies of the isomer. Overall, a model dependence 
of the results has surfaced: RMF forces tend to lower, and often too 
low, barriers and excitation energies, while most Skyrme forces tend 
to higher values, which sometimes leads to an overestimation on the 
mean-field level. 

For superheavy nuclei all models and forces agree on the systematic 
trends concerning the fission barriers, ground state deformations, 
and fission modes. There are differences in detail. Fission isomers 
are generally suppressed in superheavy elements, with the exception 
of the RMF force NL3 in our sample. For larger nuclei, there emerge 
large and systematic discrepancies between SHF and RMF concerning 
the barrier heights, reaching a factor two for \mbox{$Z=120$}. This 
amplifies and confirms the tendency which has been demonstrated for 
actinides. The factor two in fission barriers around \mbox{$Z=120$} 
means in absolute numbers that SHF fission barriers are about 5 MeV 
larger and this amounts to many orders of magnitude longer fission
lifetimes. Moreover, the SHF forces employed here predict also larger
barriers than the more phenomenological mic-mac models. The 
reason for the systematic difference between SHF and RMF have yet
to be found out. We suspect that the difference is caused by a
different shell structure in the models. This point deserves 
more investigation. 

The further systematic trends are shared by SHF and RMF. There is a
transition from well-deformed ground states around \mbox{$Z=108$} to 
nearly spherical ones at \mbox{$Z=120$}, which develops through very 
soft nuclei which might exhibit shape isomerism. And there is a marked
breakdown of fission stability around \mbox{$Z=110$} in agreement with
experimental findings where the $\alpha$ chains of superheavy
elements are limited at the lower end by fission.

The axially symmetric fission barriers are, of course, only a first
indicator of fission stability (to be more precise an upper
limit). One needs yet to include triaxial degrees of freedom and to
model the dynamics of fission to obtain life times.
Taking into account the results from actinides and superheavy nuclei
the mean-fields models used in this study seem to deliver lower 
(and probably too low - RMF) and upper (SHF) limits for the barrier
heights. It is yet difficult to map these differences on special and 
isolated features of the models which remains an urgent and important 
task for the near future.
%
%===========================================================================
%
\begin{acknowledgments} 
We thank our colleagues W.\ Greiner, Yu.~Ts.\ Oganessian, M.\ Samyn, and
V.~I.\ Zagrebaev for inspiring discussions which initiated this work
as well as S.\ Schramm for help with coding problems.
This work was supported in part by Bun\-des\-ministerium f\"ur Bildung 
und Forschung (BMBF), Project No.\ 06 ER 808, and by the PAI-P5-07 of 
the Belgian Office for Scientific Policy. M.~B.\ acknowledges support 
through a European Community Marie Curie Fellowship.
\end{acknowledgments}
%
%===========================================================================
%

\end{document}